\pdfoutput=1
\documentclass[longbibliography,aps,prl,reprint]{revtex4-2}
\usepackage{blindtext}

\usepackage{amsmath}
\usepackage{amssymb}
\usepackage{graphicx}
\usepackage[dvipsnames]{xcolor}
\numberwithin{equation}{section}
\usepackage{array}
\usepackage{mathtools}
\usepackage{dsfont}
\usepackage{mathrsfs}
\usepackage{tikz}\usetikzlibrary{matrix,fit}
\usepackage{varwidth}
\usepackage{enumerate}

\usepackage{empheq}

\usepackage{hyperref}
\hypersetup{
    colorlinks=false,
    linkbordercolor	=BlueViolet,
citebordercolor	=OliveGreen,
urlbordercolor=RoyalBlue
}
\usepackage{slashed}
\usepackage{upgreek}

\newdimen\mytextwidth
\newcommand\rem[2][cyan!40!green]{\noindent\nobreak\hfil\penalty1000\hfilneg
\mytextwidth=\linewidth\advance\mytextwidth by 2mm%
\begin{tikzpicture}[baseline=-\the\dimexpr\fontdimen22\textfont2\relax]\node[outer sep=0pt,draw=black,fill=#1,fill opacity=1,text opacity=1,rectangle,rounded corners]{\begin{varwidth}{\mytextwidth}\textcolor{white}{#2}\end{varwidth}};
\end{tikzpicture}\allowbreak%
}

\newcommand\whiterem[2][white!]{\noindent\nobreak\hfil\penalty1000\hfilneg
\mytextwidth=\linewidth\advance\mytextwidth by 2mm%
\begin{tikzpicture}[baseline=-\the\dimexpr\fontdimen22\textfont2\relax]\node[outer sep=0pt,draw=black,fill=#1,fill opacity=1,text opacity=1,rectangle,rounded corners,line width=1.5pt]{\begin{varwidth}{\mytextwidth}\textcolor{black}{#2}\end{varwidth}};
\end{tikzpicture}\allowbreak%
}

\newcommand{\dd}{\partial}

\newcommand{\CP}{\mathds{CP}}
\newcommand{\CC}{\mathds{C}}

\renewcommand{\bar}{\overline}

\newcommand{\bea}{\begin{equation}}
\newcommand{\eea}{\end{equation}}
\newcommand{\bear}{\begin{eqnarray}}
\newcommand{\eear}{\end{eqnarray}}
\newcommand{\bearr}{\begin{eqnarray*}}
\newcommand{\eearr}{\end{eqnarray*}}

\newcommand{\appendixnumberline}[1]{Appendix #1.\space}

\let\oldappendix\appendix
\makeatletter
\renewcommand{\appendix}{%
  \addtocontents{toc}{\let\protect\numberline\protect\appendixnumberline}%
  \renewcommand{\@seccntformat}[1]{\large\bfseries Appendix . }%
  \oldappendix
}
\makeatother

\usepackage{mdframed}

\DeclareFontFamily{U}{solomos}{}
\DeclareErrorFont{U}{solomos}{m}{n}{10}
\DeclareFontShape{U}{solomos}{m}{n}{
  <-> s*[1.1]  gsolomos8r
}{}

\newcommand{\vkappa}{\text{\usefont{U}{solomos}{m}{n}\symbol{'153}}}

\usepackage{amsmath}
\numberwithin{equation}{section}

\begin{document}
\title{Integrable sigma models on Riemann surfaces}
\author{Dmitri Bykov}
\email{bykov@mi-ras.ru, dmitri.v.bykov@gmail.com}
\affiliation{1) Steklov Mathematical Institute of Russian Academy of Sciences, Moscow, Russia}
\affiliation{2) Institute for Theoretical and Mathematical Physics, Lomonosov Moscow State University, Moscow, Russia}
\begin{abstract}
We consider quantum aspects of a class of generalized Gross-Neveu models, which in special cases reduce to sigma models. We show that, in the case of gauged models, an admissible gauge is $A_\mu=0$, which is a direct analogue of the conformal gauge in string models. Chiral anomalies are a gauge counterpart of the Weyl anomaly, and are required to vanish. Topological effects on the worldsheet lead to an integration over moduli spaces of connections on a Riemann surface. This is an initial step in studying the effects of worldsheet geometry and topology in integrable sigma models.
\end{abstract}

\maketitle

Recently a new approach to sigma models with complex homogeneous target spaces such as $\CP^{n-1}$, Grassmannians, flags etc. has been proposed~\cite{BykovRicci, BykovSUSY, BykovSMGN, BykovReview}. The approach, based on an exact equivalence with gauged chiral Gross-Neveu models involving both bosonic and fermionic fields, offers many calculational benefits as compared to the standard formulation, provides a new take on SUSY models and should prove useful in the analysis of quantum integrability. A somewhat analogous first-order formulation is also helpful in 4D Yang-Mills theory~\cite{CostelloSDYM}.

Among other matters, in~\cite{BykovRicci} it was observed that the one-loop beta functions of such theories are independent of the gauge fields, so that the same beta function may be shared by several inequivalent models. Motivated by this curious property, in the present letter we systematically study the role of gauge fields in these models. The key result is that they carry only topological degrees of freedom, parametrizing moduli spaces of holomorphic vector bundles over the worldsheet $\mathsf{\Sigma}$ (which is assumed a Riemann surface).  In particular, for $\mathsf{\Sigma}=\mathbb{R}^2$ one can completely eliminate the gauge field by choosing a gauge $A_\mu=0$. The setup is reminiscent of string sigma models, where the worldsheet metric may be eliminated by going to conformal gauge, and in general one has to integrate over conformal classes of metrics. Our models provide a gauge field version of that gravitational setup (also studied in~\cite{Hori} in the case of gauged WZNW models). The gauge counterpart of the Weyl anomaly, which is carefully cancelled in string models~\cite{PolyakovBosonic, TseytlinSigma}, is the chiral anomaly that is also required to vanish. The cancellation may be achieved by including fermions in various ways.

There are several facts hinting at the integrability of the proposed models in flat space. One piece of evidence is the integrability of the fermionic Gross-Neveu model~\cite{Dashen} (both classical~\cite{NeveuPapa} and quantum~\cite{Andrei, Destri}). Another one, perhaps more familiar from sigma model theory~\cite{Pohlmeyer, Zakharov}, has to do with the existence of a family of flat connections (for a review cf.~\cite{Zarembo, ArutyunovLect}). Finally, as is typical for integrable models (see~\cite{Klimcik1, Klimcik2, Delduc, Klimcik} for the background and~\cite{CYa} for the relevant modern developments), there are canonical trigonometric/elliptic deformations, and the deformed geometry is stable under RG-flow, at least at one loop~\cite{BykovRicci, BykovSMGN} (for earlier developments cf.~\cite{FOZ, Valent, Lukyanov, Litvinov, Hoare, Levine}). It was also conjectured in~\cite{BykovRicci} that the anomalies known to obstruct integrability of the bosonic $\CP^{n-1}$ model~\cite{AbdallaAnomaly, AbdallaCancel, AbdallaBook} are cancelled in models with vanishing chiral anomalies.

Throughout most of the letter we treat the $\CP^{n-1}$ model in detail, touching upon the non-Abelian case in the last section. Our results suggest that the theories of interest may be consistently placed on a Riemann surface without spoiling some of the crucial features. Whether integrability is preserved during this process is a subject of further study.

\vspace{-0.3cm}
\section{The $\CP^{n-1}$ sigma model as a Gross-Neveu model}

In~\cite{BykovRicci, BykovSUSY, BykovSMGN} we reformulated the familiar $\CP^{n-1}$ sigma model as a generalized Gross-Neveu model as follows. Consider a column vector $U\in \CC^n$ and a row vector $V\in \CC^n$, together with a complex gauge field $\mathcal{A}$, and write down the Lagrangian
\bea\label{cplagr}
\mathcal{L}=V\bar{D}U+\bar{U}D\bar{V}+\vkappa\,(\bar{U}U)(V \bar{V})\,,
\eea
where $\bar{D}U=\bar{\dd}U-i\bar{\mathcal{A}}\,U$. Here $\bar{\dd}={\dd\over \dd \bar{z}}$ is the derivative w.r.t. a holomorphic coordinate $z$ on $\mathsf{\Sigma}$ and $\mathcal{A}$ is the $(0, 1)$ component of a Hermitian (unitary) connection $A= \mathcal{A}\,dz+\bar{\mathcal{A}} \,d\bar{z}$. Grouping the $U$ and $\bar{V}$ fields in a single Dirac spinor, one recognizes in~(\ref{cplagr}) the Lagrangian of a chiral Gross-Neveu model~\cite{GrossNeveu, WittenThirring} in bosonic incarnation. Classically~(\ref{cplagr}) has a $\CC^\ast$ gauge symmetry ($\upchi\in \CC$)
\bea\label{gaugetrans}
U\to e^{\upchi} U, \;V\to e^{-\upchi} V, \;\bar{\mathcal{A}}\to\bar{\mathcal{A}}-i\,\bar{\dd} \upchi\,.
\eea
Here $\CC^\ast=U(1)\times \mathbb{R}^\ast$ is the chiral symmetry on a Riemannian worldsheet~\cite{Zumino, BykovRicci}, where $U(1)$ stands for vectorial, and $\mathbb{R}^\ast$ for axial, transformations.

One can eliminate the $V,\bar{V}$ variables from~(\ref{cplagr}) using the e.o.m., arriving at the `geometric' form of the Lagrangian
\bea
\mathcal{L}\simeq {1\over \vkappa}\,\frac{|\bar{D}U|^2}{|U|^2}\,.
\eea
Using the $\mathbb{R}^\ast$ part of the gauge symmetry to set the gauge $|U|^2=1$, one obtains the standard GLSM form of the $\CP^{n-1}$ sigma model, up to a topological term.

\section{The gauge $A=0$ as analogue of conformal gauge}

Let us first study the gauge transformations~(\ref{gaugetrans}) for the case of the simplest possible worldsheet $\mathsf{\Sigma}=\mathbb{R}^2$, assuming decay conditions for $\mathcal{A}$ at infinity. If anomalies are absent, somewhat surprisingly an admissible gauge is
\bea\label{Aeq0}
A=\mathcal{A}=\bar{\mathcal{A}}=0\,,
\eea
i.e. one can eliminate the gauge field altogether. Indeed, one can explicitly solve the equation ${\bar{\mathcal{A}}-i\,\bar{\dd} \upchi=0}$ that leads to~(\ref{Aeq0}): by the Cauchy-Green formula, $\upchi(z, \bar{z})={i\over \pi}\,\int d^2w\,{1\over z-w} \,\bar{\mathcal{A}}(w, \bar{w})$.

The gauge~(\ref{Aeq0}) is a direct analogue of conformal gauge in string models, where instead the metric may be completely eliminated. Recall that imposing the conformal gauge  involves using both $\mathrm{Diff}$- and Weyl-invariance. In the gauge system at hand the analogue of $\mathrm{Diff}$-symmetry is the usual $\mathsf{G}$-gauge invariance, where $\mathsf{G}$ is a compact reductive group (such as $\mathrm{U}(1)$), and the analogue of Weyl symmetry corresponds to axial gauge transformations related to the non-compact part $\mathsf{G}_{\CC}/\mathsf{G}$ (such as~$\mathbb{R}^\ast$). 

\vspace{-0.2cm}
\section{Chiral anomalies}

Chiral gauge transformations, crucial for imposing the gauge~(\ref{Aeq0}), are typically anomalous quantum-mechanically, and care should be taken to ensure the anomalies cancel.  Recall that vanishing of the Weyl anomaly leads to the central charge being zero. Cancellation of the chiral anomaly means that the level of the corresponding Kac-Moody algebra should vanish: $\mathsf{k}=0$. 

\vspace{-0.3cm}
\subsection{BRST quantization}

One way of arriving at the condition $\mathsf{k}=0$ is by performing BRST quantization in the gauge~(\ref{Aeq0}). Since the gauge transformation is $\delta \mathcal{A}=i \dd \bar{\chi}$, $\delta \bar{\mathcal{A}}=-i\bar{\dd} \chi$, the part of the action related to gauge-fixing has the form
\bea
\mathcal{S}_{\mathrm{gf}}=i\int\,d^2z\,\left(\,\bar{\uplambda}\, \mathcal{A}+\uplambda\, \bar{\mathcal{A}}+b\bar{\dd} c-\bar{b}\dd \bar{c}\,\right)\,,
\eea
where $\uplambda, \bar{\uplambda}$ are Lagrange multipliers and $b, c$ are the ghost fields.
This has the obvious off-shell invariance
$
\delta \bar{\mathcal{A}}=-i\varepsilon \bar{\dd} c\,,\,\, \delta b=i\varepsilon \uplambda\,.
$ 
To ensure invariance of the matter part~(\ref{cplagr}), we additionally postulate the transformation laws
$
\delta U=\varepsilon c\, U$, $\delta V=-\varepsilon c\, V\,.
$

One can eliminate the Lagrange multipliers and pass over to on-shell BRST transformations. To this end, we take the variation of the full action w.r.t. the gauge fields, which leads to $\uplambda-VU=0$, so that on-shell transformations take the form
\bea
\delta b=i\varepsilon  VU\,,\quad \delta U=\varepsilon c\, U\,,\quad \delta V=-\varepsilon c\, V\,.
\eea
The BRST current is $j_{\mathrm{BRST}}=c\,J$, where $J=VU$ is the $\mathrm{U}(1)$ chiral algebra current with OPE $J(z) J(w)=\frac{\mathsf{k}}{(z-w)^2}+\ldots$ Here $\mathsf{k}=-n$ is the level of the chiral algebra. Using nilpotency of  $c(z)$, one finds
\bea\label{BRSTope}
j_{\mathrm{BRST}}(z) j_{\mathrm{BRST}}(w)=\mathsf{k}\,\frac{c\dd c(z)}{z-w}+\ldots 
\eea

On a cylinder $\mathsf{\Sigma}=\mathbb{R}\times S^1$ one can define a BRST charge $\mathcal{Q}:=\int_{S^1}\,dz\,j_{\mathrm{BRST}}(z)$. It follows from~(\ref{BRSTope}) that $\mathcal{Q}^2=\mathsf{k}\,\sum_{j\in \mathbb{Z}}\,j c_j c_{-j}\,,$ where $c_j$ are the Fourier modes of~$c(z)$. Nilpotency of the BRST charge requires
$
\mathsf{k}=0\,,
$ 
i.e. that the central extension in the chiral algebra of the gauge group should vanish. This is equivalent to the vanishing of chiral gauge anomalies. The purely bosonic model~(\ref{cplagr}) is anomalous, since here $\mathsf{k}=-n$. Anomaly cancellation may be achieved by adding fermions, which make a positive contribution to the level. More generally, assuming $\mathfrak{h}$ is the Lie (super)algebra of the complex gauge (super)group, the cancellation condition is
\bea\label{WZWanomcancel0}
\mathrm{Str}_{\mathsf{W}}(\tau^a \tau^b)=0\quad \textrm{for}\quad \tau^a, \tau^b\in \mathfrak{h}\,,
\eea
where $\mathsf{W}$ is the representation of the matter fields (bosonic and fermionic). This reduces to $\mathsf{k}=0$ if ${\mathrm{Str}_{\mathsf{W}}(\tau^a \tau^b)=\mathsf{k}\,\delta^{ab}}$.  Condition~(\ref{WZWanomcancel0}) is a direct counterpart of the anomaly cancellation condition~\cite{WittenWZWgauging} for   WZNW models on Riemannian worldsheets (these are encountered in~\cite{Gawedzki, Nair, MaldacenaOoguri}), extended to gauge supergroups.

\vspace{-0.3cm}
\subsection{Anomaly in `lightcone' gauge} 

To complete the parallel with gravitational anomalies in string sigma models, recall that the latter manifest themselves in lightcone gauge via an anomaly in the (target space) Lorentz symmetry algebra. The same phenomenon occurs in the model~(\ref{cplagr}). Here lightcone gauge is replaced by the `inhomogeneous gauge' $U_n=1$. Generically this can be achieved by the~$\CC^\ast$ gauge symmetry. The remaining $U_i$, $i=1, \ldots n-1$ coordinates are the inhomogeneous coordinates on~$\CP^{n-1}$. Anomalies are related to the kinetic term in~(\ref{cplagr}), which is invariant under the global affine symmetry $\widehat{\mathfrak{sl}_n}$. The chosen gauge explicitly breaks it down to $\widehat{\mathfrak{gl}_{n-1}}$, and the question is whether the original symmetry is realized non-linearly.

Consider the cubic generators
\bea
\mathsf{L}_{ni}=U_i\sum\limits_{j=1}^{n-1}\,U_jV_j+\upxi\, \dd U_i\,,\quad\quad i=1, \ldots n-1\,.
\eea
The last term features a parameter $\upxi$ and represents a normal ordering ambiguity. The OPE $\mathsf{L}_{ni}(z)\mathsf{L}_{nk}(w)$ would be non-singular in the $\widehat{\mathfrak{sl}_n}$ symmetry algebra. Here instead, using the basic OPE $U_i(z) V_k(w)=\frac{\delta_{ik}}{z-w}+\ldots$ , we find
\bear\nonumber
&&\mathsf{L}_{ni}(z)\mathsf{L}_{nk}(w)=-\frac{n+2+2\upxi}{(z-w)^2}\,U_i(z) U_k(z)+\\ \nonumber&&+\frac{1}{z-w}\left((1+n+\upxi)\,U_i\dd U_k(z)+(1+\kappa)\,U_k\dd U_i(z)\right)+\ldots
\eear
Setting $\upxi=\!-1\!-{n\over 2}$ to cancel the first term, 
\bea\label{LiLkOPE}
\mathsf{L}_{ni}(z)\mathsf{L}_{nk}(w)=\frac{n}{2}\cdot \frac{U_i\dd U_k(z)-U_k\dd U_i(z)}{z-w}+\ldots
\eea
Both anomalies~(\ref{BRSTope}) and~(\ref{LiLkOPE}) are reminiscent of the expressions occurring in string sigma models~\cite{GSW, BLT}.

The anomaly~(\ref{LiLkOPE}) has been observed in~\cite{NekrasovBG, Witten02} in the context of $\beta\gamma$-systems. It vanishes in the special case $n=2$: here the indices $i=k=1$ in~(\ref{LiLkOPE}) take only one value, and the two terms cancel. This leads to the $\widehat{\mathfrak{sl}_2}$ chiral algebra at the critical level $\mathsf{k}_{\mathfrak{sl}_2}=-2$~\cite{MSV, Witten02}, where the  Sugawara energy-momentum tensor is identically zero, making a sigma model interpretation obscure.

\vspace{-0.2cm}
\subsection{The `linear axion'}

One way of cancelling the Weyl anomaly in string models is by introducing a linear dilaton. An Abelian gauge anomaly may be cancelled by a similar mechanism. Indeed, Schwinger's effective action for the gauge field in~(\ref{cplagr})~\cite{MooreVafa} has the form
\bea\label{Schwing}
\mathcal{S}_{\mathrm{eff}}=-{n\over 4\pi}\,\int\,d^2z\,F{1\over \triangle} F\,,
\eea
where $F=i(\bar{\dd}\mathcal{A}-\dd \bar{\mathcal{A}})$. This can be cancelled by an additional scalar field $\phi$ with the `linear axion' action:
\bea
\mathcal{S}_{\phi}={1\over 2\pi}\int\,d^2z\,\left({1\over 2}(\dd_\alpha \phi)^2+ n^{1\over 2}\,\phi \cdot F\right)
\eea
Elimination of $\phi$ via its e.o.m. provides a contribution equal to~(\ref{Schwing}) but with an opposite sign. Alternatively, one can fermionize $\mathcal{S}_{\phi}$~\cite{Basso} arriving at the action of a single Dirac fermion with charge $Q=n^{1\over 2}$. Therefore this is another application of the anomaly cancellation mechanism by fermions (we have $\mathrm{Str}(\tau^2)=1\times n-Q^2\times 1=0$, in line with~(\ref{WZWanomcancel0})). This type of coupling features in the model arising in a limit of the $AdS_4\times \CP^3$ superstring~\cite{BykovCP3, Basso}.

\section{Curved worldsheet}

As a next step, we wish to couple the Gross-Neveu model to a (fixed) worldsheet metric on a Riemann surface $\mathsf{\Sigma}$. The classical action $S=\int d^2z \,\mathcal{L}$ with Lagrangian~(\ref{cplagr}) then defines the theory in conformal coordinates.  For it to make sense in this extended setup, $VU\,dz$ should be a section of the canonical bundle $K_{\mathsf{\Sigma}}$. Therefore we may take $U$ and $V$ as sections of $N_{\mathrm{grav}}\otimes N_{\mathrm{gauge}}$ and $K_{\mathsf{\Sigma}}\otimes N_{\mathrm{grav}}^{-1} \otimes N_{\mathrm{gauge}}^{-1}$, respectively, where $N_{\mathrm{grav}}$ is a line bundle over $\mathsf{\Sigma}$ characterizing the spin of the matter field, and $N_{\mathrm{gauge}}$ is the line bundle corresponding to the gauge field $A$. 

\vspace{-0.4cm}
\subsection{The mixed anomaly}

Upon coupling the theory to a worldsheet metric, one should make sure that no mixed gauge-gravitational anomaly arises. Denote by $\mathcal{J}$ the matrix of integer-normalized gravitational charges (spins) characterizing the bundle $N_{\mathrm{grav}}$ for various fields ($\mathcal{J}=0$ corresponding to spin-$1/2$). We assume it commutes with the gauge generators $\tau_a$ and with any global symmetry that one wants to keep. The condition for the vanishing of the mixed anomaly reads: 
\bea\label{WZWanomcancel1}
\mathrm{Str}_{\mathsf{W}}(\mathcal{J} \tau_a)=0 \quad \textrm{for}\quad \tau^a\in \mathfrak{h}\,.
\eea
In the language of the CFT system discussed earlier, this is tantamount to requiring that the current $J(z)$ be a primary operator.

The two conditions~(\ref{WZWanomcancel0}) and~(\ref{WZWanomcancel1}) form a complete set of anomaly cancellation conditions: (\ref{WZWanomcancel0}) is a condition on the theory in flat space, whereas~(\ref{WZWanomcancel1}) restricts the ways how it could couple to the worldsheet metric. It was shown in~\cite{BykovSUSY} that in several important cases (the SUSY case and the case of `minimally coupled fermions')  $\mathrm{Str}_{\mathsf{W}}(\tau_a)=0$ holds, so that $\mathcal{J}=q\mathds{1}$ ($q\in \mathbb{Z}$). This means that bosons and fermions in $\mathsf{W}$ have the same gravity couplings, which corresponds to the A-type topological twist~\cite{WittenTopSigma, WittenABtwists, Mirror}.

\vspace{-0.2cm}
\subsection{Global aspects}

When the topology of $\mathsf{\Sigma}$ is non-trivial, the gauge~(\ref{Aeq0}) cannot be imposed. First, observe that the degree $p:={1\over 2\pi}\int_{\mathsf{\Sigma}} \,dA$ is gauge-invariant, since $\delta p={1\over 2\pi}\int_{\mathsf{\Sigma}} \,d\ast d \,\mathrm{Re}(\upchi)=0$ as integral of a total derivative (for bounded~$\upchi$). We may decompose $A=A^{(0)}+\widehat{A}$, where $A^{(0)}$ is a fixed background gauge field satisfying ${1\over 2\pi}\int_{\mathsf{\Sigma}} \,dA^{(0)}=p$, and ${1\over 2\pi}\int_{\mathsf{\Sigma}} \,d\widehat{A}=0$. Using chiral gauge transformations, we may then set $d\widehat{A}=0$~\cite{DonaldsonRiemann}. The flat connection leads to additional gauge invariants -- the holonomies $\exp{\left(i\oint_{\gamma} \widehat{A}\right)}$ for $\gamma \in \pi_1(\mathsf{\Sigma})$.

In the Abelian case these invariants parametrize the moduli space of line bundles $N$ over $\mathsf{\Sigma}$, which is
\bea\label{PicSigma}
\mathrm{Pic}(\mathsf{\Sigma})\simeq \mathbb{Z}\times \mathrm{Jac}(\mathsf{\Sigma})\,,
\eea
where $\mathbb{Z}$ corresponds to the degree of $N$ and (assuming the surface is of genus $g$) $\mathrm{Jac}(\mathsf{\Sigma})\simeq \mathsf{T}^{2g}$ is the Jacobian of $\mathsf{\Sigma}$ (cf.~\cite{MooreVafa}). Holonomies may be thought of as coordinates on $\mathrm{Jac}(\mathsf{\Sigma})$. An explicit description of elements of~(\ref{PicSigma}) involves the theory of theta functions~\cite{Mumford}.

Additional insight comes from the analysis of ghost zero modes. The zero modes of $c$ correspond to residual gauge transformations preserving the gauge $A=0$. These are holomorphic functions on $\mathsf{\Sigma}$ (constant for compact $\mathsf{\Sigma}$). Since 
$b(z)dz$ is a one-form on $\mathsf{\Sigma}$, the zero modes of~$b$ are holomorphic one-forms. On a surface of genus $g$ there are exactly~$g$ of those, and they parametrize the tangent space to $\mathrm{Jac}(\mathsf{\Sigma})$.

\subsection{$\mathsf{\Sigma}=S^2$: bundles of non-zero degree}

Let us illustrate the meaning of the discrete part $\mathbb{Z}$ in~(\ref{PicSigma}), taking the simplest case $\mathsf{\Sigma}=S^2$. Here  all holonomies are zero and a line bundle is characterized by its degree~$p\in\mathrm{Pic}(S^2)=\mathbb{Z}$. The connection is not flat for $p\neq 0$, but we can still find a suitable gauge, such as
\bea\label{gaugefieldS2}
A=i{p\over 2}\,\frac{zd\bar{z}-\bar{z}dz}{1+|z|^2}\,.
\eea
Recall that, for a sphere, $K_{\mathsf{\Sigma}}=\mathcal{O}(-2)$. Assuming the background bundle is $N_{\mathrm{grav}}=\mathcal{O}(q)$, we find the following gauge transformations for the fields under the change of variables $z\to z^{-1}$ (here $\varphi=\mathrm{arg}(z)$):
\bea
\label{qbundle}
U\to z^{-q} e^{-ip \varphi} \,U, \quad V\to z^{2+q} e^{ip\varphi} \,V\,.
\eea
Although the gauge field~(\ref{gaugefieldS2}) is not flat for ${p\neq 0}$, one finds $i\bar{\mathcal{A}}=
-{p\over 2}\,\bar{\dd}\log{(1+|z|^2)}$, so that each component can be `gauged away' by
$ U\to (1+|z|^2)^{-{p\over 2}}\,U,\, V\to (1+|z|^2)^{p\over 2}\,V$. 
In the new variables the gauge transformations are purely holomorphic:
\bea\label{pqbundle}
U\to z^{-p-q}U,\quad\quad V\to z^{2+p+q}\,V\,.
\eea
As a result, $U$ and $V$ are sections of $\mathcal{O}(p+q)$ and $\mathcal{O}(-2\!-p\!-q)$, respectively. It follows that, once all gauge line bundles $N_{\mathrm{gauge}}$ are included, there is no invariant meaning of $N_{\mathrm{grav}}$: summing over gauge bundles is the same as summing over all spins.

Quantum mechanically one should take into account the anomaly cancellation conditions. To give an example where they play a role, recall that the central charge of the free CFT system is $\mathsf{c}=\mathrm{Str}_{\mathsf{W}}\left(3\,\mathcal{J}^2-\mathds{1}\right)$. Now, consider the change of the gravitational couplings by elements from the center of $\mathfrak{h}$ ($s_a$ are constants):
$\mathcal{J}\to\mathcal{J}+\sum_{\tau_a\in\mathcal{Z}(\mathfrak{h})}\,s_a\,\tau_a$. This is a generalization of the shift from $N_{\mathrm{grav}}=\mathcal{O}(q)$ to $N_{\mathrm{grav}}=\mathcal{O}(p+q)$ in~(\ref{qbundle})-(\ref{pqbundle}). One sees that the conditions~(\ref{WZWanomcancel0}), (\ref{WZWanomcancel1}) ensure invariance of $\mathsf{c}$ under this shift. 

Finally, bundles of non-zero degree $p$ exist on arbitrary Riemann surfaces. In that case one can take $A={i p\over 2}\,\dd\Omega \,dz-{i p\over 2}\,\bar{\dd}\Omega \,d\bar{z}$ for the connection ($\Omega$ being the `potential'). One can again get rid of the gauge field by the transformation ${U\to e^{p\Omega} U}$, $V\to e^{-p\Omega} U$ and assume that $U$ and $V$ are sections of the relevant line bundles. Similar questions have been discussed in the context of the quantum Hall effect on Riemann surfaces, cf.~\cite{HaldaneQHE, Iengo, Klevtsov}.

In the following sections we will analyze the role of the Jacobian in~(\ref{PicSigma}). As a first example we take up the case when the worldsheet is a cylinder.

\vspace{-0.2cm}
\subsection{$\mathsf{\Sigma}=\mathds{R}\times S^1$: the Hilbert space interpretation}

On a cylinder $\mathsf{\Sigma}=\mathds{R}\times S^1$ we regard $\mathds{R}$ as the `space' direction and~$S^1$ as the compactified Euclidean time direction. We assume decay conditions for the curvature $F$ at infinity, so that the integrals $\oint_{S^1}\,A\big|_{\pm \infty}$ are independent of the contours. The gauge invariants are the `degree' $p=\!{1\over 2\pi}\!\left(\oint_{S^1} A\big|_{+ \infty}\!\!\!-\oint_{S^1} A\big|_{- \infty}\right)\in\mathbb{R}$ and the holonomy $h:= \exp{\left(i \oint_{S^1}\,A\big|_{+ \infty}\right)}\in \mathrm{U}(1)$. A particularly simple interpretation exists for the sector $p=0$. In this case we may eliminate the gauge field at the expense of imposing $h$-twisted boundary conditions along the circle~$S^1$ (cf.~\cite{BykovReview}). This leads to the following expression for the partition function in an external gauge field~$A$: $\mathcal{Z}_{p=0}(A)=\mathrm{Tr}\left(h\,e^{-\beta H}\right)$, where $H$ is the Hamiltonian of the theory on $\mathds{R}$. Integrating over the gauge field is tantamount to averaging over the twist:
\bea\label{twistint}
\mathcal{Z}_{p=0}=\int dh\,\mathrm{Tr}\left(h\,e^{-\beta H}\right)=\mathrm{Tr}_{ \mathrm{inv}}\left(e^{-\beta H}\right)\,,
\eea
i.e. the Hilbert space is projected to the subspace of states invariant w.r.t. the gauge group. 

\section{Non-Abelian gauge groups}

The  simple case of the $\CP^{n-1}$ model has a generalization~\cite{BykovSMGN} related to quiver varieties~\cite{Nakajima, NakajimaBook}. The gauge groups would then generally be non-Abelian -- an example is provided by the Grassmannian $\mathrm{Gr}_{k, n}$ target space with gauge group $\mathrm{U}(k)$. The gauge transformation law~(\ref{gaugetrans}) is then replaced by
\bea\label{complgaugetransfo}
\bar{\mathcal{A}}\to g\bar{\mathcal{A}} g^{-1}-i\,\bar{\dd}g \,g^{-1}\,,\quad g\in \mathrm{GL}(k, \CC)\,.
\eea
Equivalence classes of connections w.r.t. these transformations define equivalence classes of rank-$k$ holomorphic vector bundles over $\mathsf{\Sigma}$. As is typical for complex quotients, there is an alternative description as a symplectic quotient, if one considers the natural symplectic form $\omega=\int_{\mathsf{\Sigma}} \mathrm{Tr}\left(\delta A \wedge \delta A\right)$ (see~\cite{AtiyahBott, Jeffrey} for a review). The corresponding moment map equation is
\bea\label{centconn}
dA-i\,A\wedge A={p\over k}\,\mathrm{vol}_{\mathsf{\Sigma}}\,\mathds{1}_n\,,
\eea
where $\mathrm{vol}_{\mathsf{\Sigma}}$ is the $2\pi$-normalized volume form on~$\mathsf{\Sigma}$. One may think of~(\ref{centconn})  as a partial gauge for the $\mathrm{GL}(k, \CC)$ gauge symmetry~(\ref{complgaugetransfo}), whose residual gauge transformations are the (standard) unitary ones acting on $A$.

The r.h.s. of~(\ref{centconn}) involves a central `Fayet-Iliopoulos' term, and the coefficient $p\over k$ is chosen so that the integral ${1\over 2\pi}\int\, \mathrm{Tr}(F)=p$ is equal to the degree of the bundle. For fixed $p$ the space of solutions to~(\ref{centconn}) may be 
described via representations of the central extension of $\pi_1(\mathsf{\Sigma})$~\cite{Narasimhan, AtiyahBott} (again assuming $g$ is the genus):
\bea\label{PSUnmoduli}
\mathcal{M}^{(p)}
=\left\{\prod\limits_{i=1}^g\,
\mathsf{A}_i \mathsf{B}_i \mathsf{A}_i^{-1} \mathsf{B}_i^{-1}=e^{2\pi i p\over k}\right\}/\mathrm{U}(k)\,,
\eea
Here $\mathsf{A}_i, \mathsf{B}_i$ are $\mathrm{U}(k)$-valued matrices representing holonomies $\mathrm{P}\exp{\left( i\oint A\right)}$ along the $\mathsf{A}$- and $\mathsf{B}$-cycles of~$\mathsf{\Sigma}$. We may as well get rid of the non-Abelian part of the gauge connection at the expense of imposing twisted boundary conditions: for example, one has $U\to \mathsf{A}_i\,U$ as one moves around the $\mathsf{A}$-cycle on $\mathsf{\Sigma}$. The remaining Abelian part may be included in the spin of the fields, as in the $\mathsf{\Sigma}=S^2$ case studied above.

Calculating the partition function of the theory should therefore proceed in the following steps:
\begin{itemize}
\item For `instanton number' $p\in \mathbb{Z}$, calculating the partition function with fixed twists along each cycle. 
\item Integrating over the moduli space~$\mathcal{M}^{(p)}$ of such twists (the torus $\mathrm{Jac}(\mathsf{\Sigma})=\mathsf{T}^{2g}$ in the Abelian case) .
\item \vspace{-0.3cm} Summing over all `instanton numbers' $p$.
\end{itemize}
This is the generalization to an arbitrary Riemann surface of the procedure encountered in~(\ref{twistint}) in the case of a cylinder.

Calculation of the partition function with generic twists seems a formidable task, but one can hope that flat-space integrability  would leave a sufficiently strong imprint on the theory on $\mathsf{\Sigma}$ to make this possible. In this case the study of integrability would again lead to flat (Lax) connections. We leave it as an open problem for the future.

\vspace{0.4cm}
\textbf{Acknowledgments.}
I would like to thank A.~A.~Tseytlin for valuable discussions and comments on the manuscript, and A.~A.~Slavnov for long-term support.

\vspace{-0.1cm}
\bibliography{RiemSurf} 

\end{document}